\documentclass[twocolumn,amsmath,amssymb,aps,preprintnumbers,showpacs]{revtex4}

\usepackage{ulem}
\usepackage{graphics}
\usepackage{dcolumn}
\usepackage{bm}
\usepackage{amsmath}
\usepackage{amssymb}
\usepackage[ps2pdf,hyperindex=true,final=true,colorlinks=true,pagebackref=false,bookmarks=true,bookmarksopen=true,bookmarksnumbered=true]{hyperref}
\usepackage{epsfig}

\RequirePackage{xspace}

\newcommand{\ie}{\textit{i.e.}\xspace}
\newcommand{\eg}{\textit{e.g.}\xspace}

\newif\iffig
\newif\ifgraph

\figtrue                     %

\graphtrue                   %

\begin{document}

\title{Magnetic hysteresis effects in superconducting coplanar microwave resonators}

\author{D.~Bothner}\email{daniel.bothner@uni-tuebingen.de}
\author{T.~Gaber}
\author{M.~Kemmler}
\author{D.~Koelle}
\author{R.~Kleiner}
\affiliation{Physikalisches Institut -- Experimentalphysik II and Center for Collective Quantum Phenomena in LISA$^+$, Universit\"{a}t T\"{u}bingen, Auf der Morgenstelle 14, 72076 T\"{u}bingen, Germany}

\author{S.~W\"{u}nsch}
\author{M.~Siegel}
\affiliation{Institut f\"{u}r Mikro- und Nanoelektronische Systeme, Karlsruher Institut f\"{u}r Technologie, Hertzstrasse 16, 76187 Karlsruhe, Germany}

\date{\today}

\begin{abstract}

We performed transmission spectroscopy experiments on coplanar half wavelength niobium resonators at a temperature $T=4.2\,$K.
We observe not only a strong dependence of the quality factor $Q$ and the resonance frequency $f_{\rm res}$ on an externally applied magnetic field but also on the magnetic history of our resonators, \ie on the spatial distribution of trapped Abrikosov vortices in the device. 
We find these results to be valid for a broad range of frequencies and angles between the resonator plane and the magnetic field direction as well as for resonators with and without antidots near the edges of the center conductor and the ground planes.
In a detailed analysis we show, that characteristic features of the experimental data can only be reproduced in calculations, if a highly inhomogeneous rf-current density and a flux density gradient with maxima at the edges of the superconductor is assumed.
We furthermore demonstrate, that the hysteretic behaviour of the resonator properties can be used to considerably reduce the vortex induced losses and to fine-tune the resonance frequency by the proper way of cycling to a desired magnetic field.

\end{abstract}


\maketitle

\section{Introduction}

\label{sec:Introduction}
When Charles P. Bean introduced his famous model for the magnetization of hard superconductors in 1962 \cite{Bean62, Bean64}, he probably has not foreseen, that once related magnetic history effects may be of importance for circuit quantum electrodynamics \cite{Wallraff04, Hofheinz09, Niemczyk10}, quantum information processing \cite{DiCarlo09} or single particle detection \cite{Day03}.
However, in this manuscript we will demonstrate, that by taking advantage of magnetic hysteresis it is possible to tune the properties of coplanar superconducting microwave cavities as currently intensely investigated and used in the above mentioned research areas.
For many applications, the quality factor $Q$ of the resonator, which defines the photon lifetime in the cavity and the sharpness of the resonance, is demanded to be rather high.
Thus there are many current efforts to identify and suppress the different loss mechanisms \cite{Wang09, Macha10, Barends10, Lindstroem09}.
Recently advanced hybrid systems have been proposed \cite{Rabl06, Imamoglu09, Verdu09, Henschel10, Bushev11}, which couple ultracold atoms, molecules or electrons to superconducting microwave cavities or investigate the combination of artificial atoms based on superconductors with real atoms.
The experimental realization of these proposals is expected to add another loss channel to the superconducting microwave cavities, as the magnetic fields required for trapping and manipulation of the atomic systems \cite{Fortagh05, Bushev08} will lead to the presence of energy dissipating Abrikosov vortices \cite{Song09}.
Lately, there were different approaches to reduce the vortex associated energy losses in special experimental situations.
In some experiments the magnetic field can be applied parallel to the plane of the superconducting film, which reduces the flux in typical coplanar resonators by orders of magnitude.
This approach was used in experiments with spin ensembles that were coupled to microwave photons in superconducting transmission line cavities \cite{Schuster10, Kubo10}.
For experiments that require out-of-plane magnetic fields it was demonstrated, that losses due to vortices can effectively be reduced by trapping and pinning the flux lines either in a slot in the center of the resonator \cite{Song09a}, or in antidots that are patterned at the resonator edges \cite{Bothner11} or all over the superconducting chip \cite{Bothner12}.
Patterning with antidots is particularly suitable for zero field cooling experimental conditions, when the vortices enter the superconductor from the edges and form a flux density gradient, called the Bean critical state.
Under zero field cooling conditions there is also a number of hysteresis effects, which on the mesoscopic scale are related to the spatial distribution of Abrikosov vortices \cite{Jooss02, Lahl02}.
%

%
In this paper we will demonstrate that the characteristic properties of superconducting resonators (with and without antidots), \ie the quality factor $Q$, the loss factor $1/Q$ and the resonance frequency $f_{\rm{res}}$, can be tuned by taking advantage of spatial vortex redistribution.
In particular we will show, that for fixed values of a perpendicularly applied magnetic flux density $B$ the resonator losses due to vortices, $1/Q_v(B)$, can be reduced and the resonance frequency $f_{\rm{res}}(B)$ can be tuned by magnetic history.
In a more detailed analysis we find strong indications, that essential features of the measured hysteresis effects can only be described by assuming a highly inhomogeneous rf-current distribution in combination with field penetration models for thin films as first mentioned by Norris\cite{Norris70} and later discussed by Brandt and Indenbom (NBI model) \cite{Brandt93}.
The paper is organized as follows.
After this introductory we first describe the sample fabrication and characterization techniques in Sec.~\ref{sec:Fabrication}.
Then, in Sec.~\ref{sec:Hysteresis} we present and discuss our experimental data that show a hysteretic behaviour in the vortex associated energy losses and the resonance frequency in perpendicular magnetic fields.
In Sec.~\ref{sec:Model} we develop a simple model to describe the dependence of the vortex associated losses on the rf-current and vortex distribution in the resonator and compare our measurements with numerical calculations.
In Sec.~\ref{sec:Demagnetization} we discuss the possibility of exploiting the hysteresis to improve and tune the properties of the resonator in a specific magnetic field.
Hysteresis effects at higher modes of the resonator and for non-perpendicular orientations between resonator plane and magnetic field are presented and discussed in Sec.~\ref{sec:Angles} . 
Finally, Sec.~\ref{sec:Conclusion} concludes the paper.

\section{Resonator fabrication and characterization}

\label{sec:Fabrication}
\begin{figure}[h]
\centering {\includegraphics{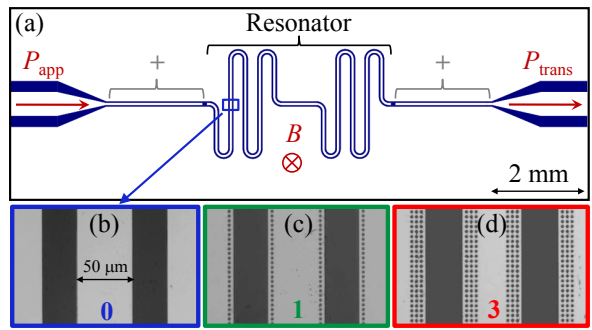}}
\caption{(Color online) (a) Layout of a $12\times4$\,mm$^2$ chip with a capacitively coupled 3.3 GHz transmission line resonator and optical images of resonators with (b) \textbf{0}, (c) \textbf{1}, and (d) \textbf{3} rows of antidots. The parts of the feedlines, which are perforated with antidots are marked with \textbf{+} (cf. Ref. \cite{Bothner11}).}
\label{fig:Figure1}
\end{figure}

We fabricated half wavelength coplanar transmission line resonators with a designed resonance frequency of $f_{\rm{res}}= 3.3$\,GHz, that are capacitively and symmetrically coupled to feed lines via $90$\,$\mu$m wide gaps at both ends.
Due to the small coupling capacitance, the resonators are strongly undercoupled\cite{Hammer07}.
In a previous study we investigated the influence of different spatial antidot arrangements on the performance of the resonators in magnetic fields and thus patterned several resonators with and without antidots\cite{Bothner11}.
In the present study we investigate these samples focusing on the magnetic hysteresis and its impact on the resonator properties.
Figure~\ref{fig:Figure1} (a) shows a sketch of the resonator layout together with optical images of the resonators without (b), with one row (c), and with three rows (d) of antidots localized at the edges of the center conductor and the ground planes.
The resonators are correspondingly named Res \textbf{0}, Res \textbf{1+} and Res \textbf{3+} with a \textbf{+} indicating that the feed lines except for the rf launch pads are also perforated with antidots, cf. Fig.~\ref{fig:Figure1} (a).
The antidots have a radius of $R=1\,\mu$m and an antidot-antidot distance of $D=4\,\mu$m.
For further design considerations of the antidots regarding size and arrangement, see Ref. \cite{Bothner11}.

All structures were fabricated on a single $330\,\mu$m thick 2 inch sapphire wafer (r-cut) by optical lithography and subsequent dc magnetron sputtering of a $d=300\,$nm thick Nb film.
After the sputtering, the wafer was cut into chips of $12\times4$\,mm$^2$, each one containing a single resonator. Finally the surplus Nb was lifted-off with acetone. 
The transmission line has a characteristic impedance of $Z_{0}\approx 54\,\Omega$, where the width of the center conductor is $S=50\,\mu$m and the gap to the ground plane is $W=30\,\mu$m.
The Nb film has a critical temperature of $T_c\approx 9\,$K and a residual resistance ratio of $R({300\,\rm{K}})/R({10\,\rm{K}})\approx 3.6$.
Each chip was mounted in a small brass box and the transmission line was electrically connected to sub-Miniature A (SMA) stripline connectors using Indium as contact material.
All measurements were performed at $T=4.2\,{\rm K}$ in liquid Helium (Helium-gas in the angle-dependent measurements).
A magnetic field perpendicular to the resonator plane could be applied with a pair of Helmholtz coils. 
We estimate the flux density seen by the resonator to be one order of magnitude larger than the applied external field due to flux focusing effects.
As we are interested in the resonator properties at magnetic fields of some mT, no measures were taken to shield the samples from earth magnetic field.
Therefore, in all zero field cooling experiments we performed, some residual field was still present, but much smaller than any field we applied (except in the angle-dependent measurements).
To characterize the resonators we applied a microwave signal of $P_{\rm{app}}=-20\,{\rm dBm}$ and frequency $f$ to one of the feed lines and measured the frequency dependent magnitude of the transmitted power $P_{\rm{trans}}(f)$ with a spectrum analyzer.
No attenuators or amplifiers were used in the measurements. We estimate the effective power at the resonator input to be about $5$--$10\,$dB lower than at the generator output, to which all values of $P_{\rm{app}}$ refer.
Similarly, the power at the input of the spectrum analyzer $P_{\rm{trans}}$ is about $5$-$10\,$dB smaller than the power directly at the resonator output.

\section{Hysteresis effects}

\label{sec:Hysteresis}
\begin{figure}[h]
\centering {\includegraphics{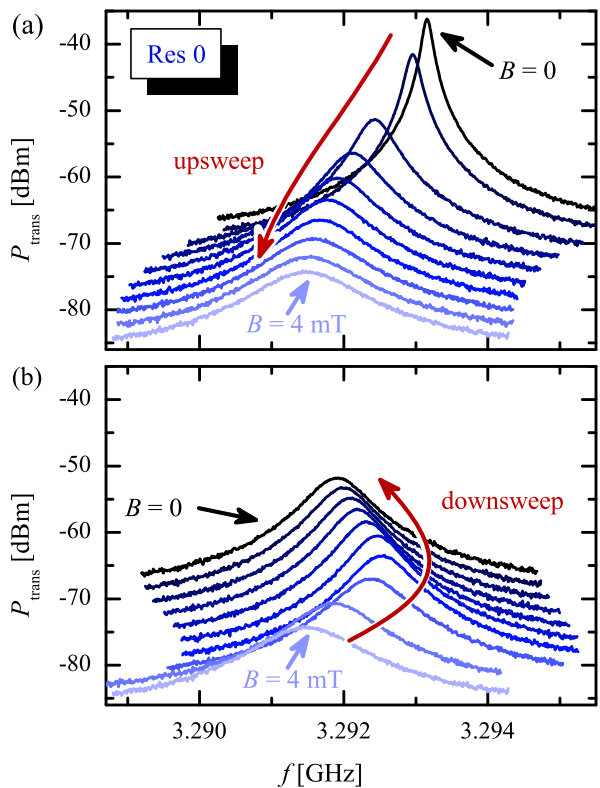}}
\caption{(Color online) Measured transmitted power $P_{\rm{trans}}$ vs. frequency $f$ of resonator $\bf{0}$ after zero field cooling, when (a) the applied magnetic field was swept from $0$ to $4\,{\rm mT}$ and (b) back to zero. Adjacent curves are shifted by $+2\,{\rm dBm}$ for better visibility, with $P_{\rm trans}(4\,{\rm mT})$ being the unshifted reference. $\Delta B\approx 0.48\,{\rm mT}$, for details see text. The applied microwave power was $P_{\rm app}=-20\,$dBm.}
\label{fig:Figure2}
\end{figure}

\begin{figure}
\centering {\scalebox{1.0}[1.0]{\includegraphics{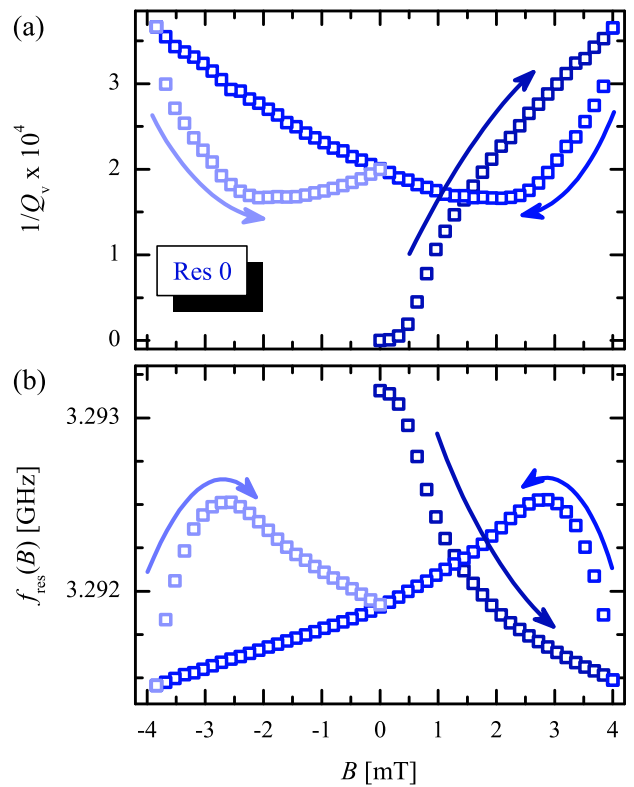}}}
\caption{(Color online) (a) Vortex associated energy losses $1/Q_v(B)$ and (b) resonance frequency $f_{\rm{res}}(B)$ of the fundamental mode $n=1$ of resonator $\bf{0}$ for a full cycle of $B$.}
\label{fig:Figure3}
\end{figure}

Figure~\ref{fig:Figure2} (a) shows the frequency dependent transmitted power $P_{\rm{trans}}(f)$ around the fundamental mode $n=1$ of a resonator without antidots (Res \textbf{0}) for different values of applied magnetic field between $B=0$ and $B=4\,$mT. 
Measurements shown were performed directly after zero field cooling of the sample.
The step size of the applied magnetic field between the different curves is $\Delta B=0.48\,$mT starting from $B=0$ with an additional curve at $B=4\,$mT.
For better visibility adjacent curves are shifted by $+2\,{\rm dBm}$, where $P_{\rm trans}(4\,{\rm mT})$ is the unshifted reference.
As has already been reported before, we find that with increasing magnetic field the resonance frequency shifts downwards and the resonance peak gets smaller and broader, indicating increasing losses\cite{Song09, Song09a, Bothner11}.
When we reduce the magnetic field from $B=4\,$mT back to $B=0$, we find a strong hysteresis in the resonance characteristics.
Figure \ref{fig:Figure2} (b) depicts the corresponding spectra (for the same values of applied magnetic field as in (a)), showing that the original state of (a) is not restored at $B=0$. Interestingly, $\partial f_{\rm res}/\partial B$ changes sign at $B\approx 3\,{\rm mT}$: After a first increase of the resonance frequency with decreasing $B$, $f_{\rm res}$ decreases again for $B\rightarrow 0$. 
We fitted the measured transmission spectra with a Lorentzian and extract the resonance frequency $f_{\rm{res}}(B)$ and the full width at half maximum $\Delta f(B)$ to quantitatively analyze the hysteresis in the resonator properties.
Using $f_{\rm{res}}(B)$ and $\Delta f(B)$ we calculate the magnetic field dependent quality factor $Q(B)=f_{\rm{res}}(B)/\Delta f(B)$ and the magnetic field dependent losses per cycle $1/Q(B)$, respectively.
In order to quantify the losses associated with the magnetic field, \ie the presence of Abrikosov vortices, we define $1/Q_v(B)\equiv 1/Q(B)-1/Q(0)$, cf. Refs. \cite{Song09, Song09a, Bothner11}.
Figure~\ref{fig:Figure3} shows the (a) vortex associated losses $1/Q_v(B)$ and (b) the resonance frequency $f_{\rm{res}}(B)$ for a full magnetic field cycle, that is $B=0\,$mT $\rightarrow B = 4\,$mT $ \rightarrow B=-3.96\,$mT $ \rightarrow B=0$.
Arrows indicate the sweep direction of the field.
Both the vortex associated losses $1/Q_v(B)$ and the resonance frequency $f_{\textrm{res}}(B)$ show a pronounced hysteretic behaviour.
Interestingly, immediately after the sweep direction is changed at $B=4\,$mT, the losses decrease considerably and are significantly smaller than the losses for the same field values, when coming from the virgin state.
In the following we indicate properties, that refer to the upsweep of $B$ with $\uparrow$ and to the downsweep with $\downarrow$. To emphasize the very first upsweep from the virgin state we use $\Uparrow$.
Over a considerable range of applied magnetic field $1/Q^{\downarrow}_v(B)<1/Q^{\Uparrow}_v(B)$ with a minimum of the losses at $B \approx 2\,$mT.
At $B\approx1.4\,$mT the loss curves from upsweep and downsweep cross and for even smaller magnetic fields $1/Q^{\downarrow}_v(B)$ increases again.
For negative magnetic field values we find a similar behaviour, where $1/Q^{\uparrow}_v(B)<1/Q^{\downarrow}_v(B)$. When we repeatedly sweep the magnetic field between $\pm B_{\rm max}=4\,{\rm mT}$, $1/Q_v(B)$ follows a butterfly like curve and never returns to the virgin state.
In view of Abrikosov vortices an obvious interpretation of this hysteresis is, that we do not get the sample vortex free again during the field cycle.
%

%
The resonance frequency in Fig.~\ref{fig:Figure3} (b) shows a hysteretic behaviour very similar to the losses but inverted regarding absolute values.
Immediately after the sweep direction is inverted, $f_{\rm{res}}$ increases and after reaching a local maximum it decreases again with further decreasing magnetic field.
Note, the magnetic field values at the downsweep, where $1/Q_v(B)$ reaches a minimum and $f_{\rm{res}}(B)$ reaches a maximum are not the same.
This discrepancy was observed for all resonators.
We believe, that $1/Q_v(B)$ and $f_{\rm{res}}(B)$ have their extrema at different field values, because the frequency shift due to a change in the kinetic inductance has not only contributions from the normal conducting vortex cores but also from the global and local screening currents, which have a slightly different distribution than the vortex cores.
%

%
Remanent vortices in type-II superconductors have already been observed and investigated in various studies before, cf. \cite{Jooss02} and references therein, leading also to hysteresis effects in the microwave properties of superconducting structures\cite{Lahl03, Bonura06, Bonura06a}.
In the following we will show, that by taking a closer look at the measured hysteresis curves one can gain a new insight into the underlying physics, as the particular shape of the curve is intimately related to the microwave current distribution as well as the vortex distribution in the resonator.

\section{The resonator loss model}

\label{sec:Model}
\begin{figure}
\centering {\scalebox{1}[1]{\includegraphics{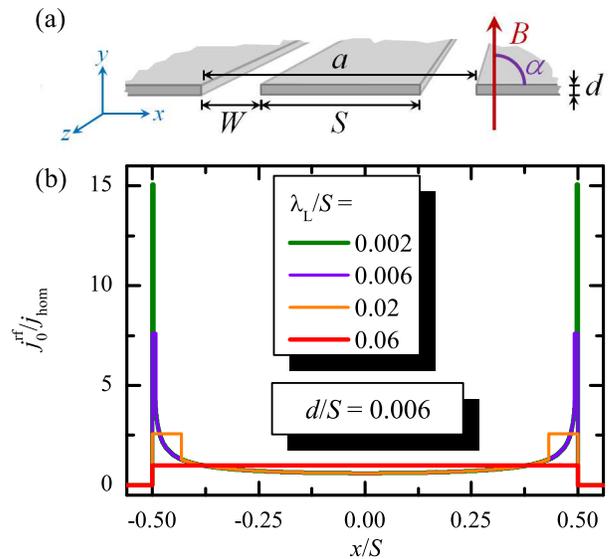}}}
\caption{(Color online) (a) Sketch of a coplanar waveguide with center conductor width $S$, ground-to-ground distance $a$ and thickness $d$; (b) Microwave current density distribution $j^{\textrm{rf}}_0(x)$ normalized to the homogeneous distribution $j_{\rm hom}=I/S$, where $I$ is the total current on the center conductor of a coplanar waveguide for different values of $\lambda_{\textrm{L}}/S$ according to Eqs.~\eqref{jrf1}, \eqref{jrf2}.}
\label{fig:Figure4}
\end{figure}

In this section we introduce a simplified model, that allows us to derive an approximate expression for the dependence of the vortex associated losses $1/Q_v$ on a spatially varying flux density $B(x)$ as well as on the microwave current density $j^{\textrm{rf}}(x)$ in a superconducting coplanar resonator.
For simplicity we only consider the center conductor of our coplanar line and ignore the ground planes.
Figure \ref{fig:Figure4} shows a sketch of a coplanar waveguide with a center conductor of width $S$ and thickness $d$.
%
In the discussion in this section the magnetic field is always applied in the y-direction and the microwave current density always points in the z-direction.
We start with classical mechanics and treat a vortex in the superconducting strip as a massless point-like particle under the influence of a driving Lorentz force $f_L=j^{\textrm{rf}}\Phi_0$ with the sheet current density $j^{\textrm{rf}}=j^{\textrm{rf}}_0\sin(\omega t)$ and a friction $f_F=\eta v$, which leads to the one-dimensional equation of motion $\eta v=j^{\textrm{rf}}\Phi_0$.
In this picture the amplitude of the vortex velocity $v$ is directly proportional to the amplitude of the alternating driving force and we find the energy dissipated per cycle $\Delta E$ to obey the proportionality $\Delta E=\int_{0}^{T}f_Lv\textrm{d}t \propto (j^{\textrm{rf}}_0)^2$ with $T=2\pi/\omega$.
If we have several vortices at different positions $x_i$ with local current densities $j^{\textrm{rf}}_0(x_i)$ we must sum up over all $\Delta E(x_i)$ from single vortices to get the overall dissipation $\Delta E_{\textrm{total}}=\sum_{i}\Delta E(x_i)\propto\sum_{i}j^{\textrm{rf}}_0(x_i)^2$.
The flux density $B(x)$ (divided by the flux quantum $\Phi_0=2.07\times10^{-15}\,$Tm$^2$) can approximately be treated as a continuous representation of the vortex density in the superconductor and with this we can rewrite the total dissipated energy per cycle and per unit length in y-direction as
\begin{equation}
\Delta E_{\textrm{total}}\propto\int_{-S/2}^{S/2}|B(x)|(j^{\textrm{rf}}_0(x))^2\textrm{d}x:=\delta e(B),
\label{Eq:deltae}
\end{equation}
where $|B(x)|$ reflects the independence of the dissipation contribution from the vortex polarity.
The herewith defined quantity $\delta e(B)$ contains all information on the spatial distributions of vortices and driving forces and is hence feasible for the qualitative study of the experimentally found hysteresis curves, although it is not the same as $1/Q_v(B)$.
However, within our approach, there is the direct proportionality $1/Q_v(B)\propto \delta e(B)$ and the proportionality factors are eliminated below as we only consider the normalized quantity $\delta e(B)/\delta e_{\textrm{max}}$.
Another possibility to find the same proportionalities between $\delta e$, $j^{\textrm{rf}}(x)$ and $B(x)$ from a more electrotechnical point of view starts with the dissipation power density $p_d(x)=\textrm{Re}[\rho_v(x)]j^{\textrm{rf}}_0(x)^2$, which must be integrated over the width $S$ of the cross section to get the overall dissipation per unit length $D=\int_{-S/2}^{S/2}p_d(x)\textrm{d}x$.
The hereby used real part of the vortex resistivity $\rho_v(x)$ can be obtained from the formulae given by different authors in different models \cite{Gittleman68, Brandt91, Coffey91}.
All of these models have the linear proportionality $\rho_v(x)\propto B(x)$ in common, what again leads us to the above expression for $\delta e(B)$.
This argument is closely related to the one, which can be found in \cite{Song09} to obtain an expression for $1/Q_v(B)$.
Interestingly, the expressions in these models are derived treating the vortices as a vortex crystallite and not as individual particles and by adding a pinning potential.
However, these assumptions do not change the proportionalities between $1/Q_v$, $j^{\textrm{rf}}(x)$ and $B(x)$.
According to Ref.~\cite{Lahl02} and references therein the microwave current density distribution $j^{\textrm{rf}}_0(x)$ in the center conductor of a coplanar line can be approximated by
\begin{equation}
\label{jrf1}
j^{\textrm{rf}}_0(x)=\frac{I}{K\left(\frac{S}{a}\right)S\sqrt{\zeta(x)}}
\end{equation}
where
\begin{equation}
\label{jrf2}
\zeta(x)=\left\{
\begin{array}{lr}
\frac{\lambda_{\textrm{eff}}}{S}\left[1-\left(\frac{S}{a}\right)^2\right], &0\le \frac{S}{2}-|x|<\lambda_{\rm eff},\\
\left[1-\left(\frac{2x}{S}\right)^2\right]\left[1-\left(\frac{2x}{a}\right)^2\right],&|x|\le \frac{S}{2}-\lambda_{\rm eff}.\\
\end{array}\right.
\end{equation}
Here, $a=S+2W$ denotes the distance between the ground planes; in our case $a=110\,\mu$m. $K$ is the complete elliptic integral and $I$ is the total current.
$\lambda_{\textrm{eff}}=\lambda_{\textrm{L}}\coth(d/\lambda_{\textrm{L}})$ is the effective penetration depth and $\lambda_{\textrm{L}}$ is the London penetration depth \cite{Klein90, Gubin05}.
For $\lambda_{\textrm{eff}}/S\ll1$, the current distribution (\ref{jrf1}) is very inhomogeneous and has pronounced maxima at the edges of the strip.
However, with increasing $\lambda_{\textrm{eff}}/S$ the maxima continuously decrease until $j^{\textrm{rf}}_0(x)$ becomes completely homogeneous, \ie $j^{\textrm{rf}}_0(x)=I/S=const.$, for $\lambda_{\textrm{eff}}\geq S/2$. 
Using the parameters of our samples we find $\lambda_{\textrm{eff}}\approx \lambda_{\textrm{L}}=100\,$nm and an approximate ratio of effective penetration depth to width of the center conductor of $\lambda_{\textrm{eff}}/S=0.002$.
Figure~\ref{fig:Figure4} (b) shows the current distribution according to Eq.~(\ref{jrf1}) for four different ratios $\lambda_{\textrm{L}}/S=0.002, 0.006, 0.02$ and $0.06$.
Note, we chose $S$ and $d$ in correspondence to our sample parameters and varied $\lambda_{\rm L}$ instead.
Therefore, $\lambda_{\rm eff}$ is different for each curve with  $\lambda_{\textrm{eff}}/S\approx0.002, 0.008, 0.069$ and $0.602$.
\begin{figure}
\centering {\scalebox{1}[1]{\includegraphics{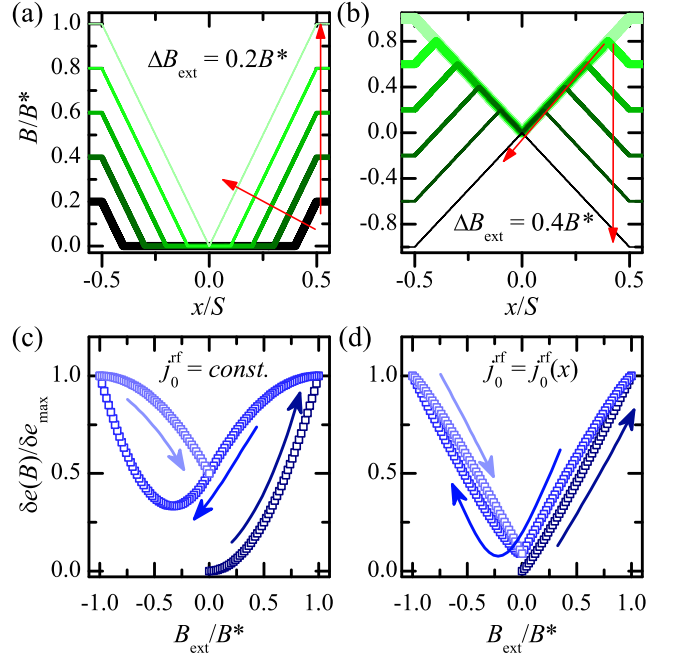}}}
\caption{(Color online) Classical Bean model flux density $B/B^*$ in a superconducting strip of width $S$ during (a) the upsweep to $B_{\textrm{ext}}=B^*$ ($B_{\textrm{ext}}/B^*=0.2, 0.4, 0.6, 0.8$ and $1$) and (b) during downsweep from $B_{\textrm{ext}}=B^*$ to $B_{\textrm{ext}}=-B^*$ ($B_{\textrm{ext}}/B^*=1, 0.6, 0.2, -0.2, -0.6$ and $-1$); Calculated $\delta e(B)/\delta e_{\textrm{max}}$ during a magnetic field cycle $B_{\textrm{ext}}/B^*=0 \rightarrow B_{\textrm{ext}}/B^*=1 \rightarrow B_{\textrm{ext}}/B^*=-1 \rightarrow B_{\textrm{ext}}/B^*=0$ assuming the classical Bean flux density and (c) a homogeneous as well as (d) a highly inhomogeneous microwave current density with $\lambda_L/S=0.002$ and $d/S=0.006$. Arrows in (a, b) indicate the progression of the subsequent flux profiles, arrows in (c, d) indicate the sweep direction.}
\label{fig:Figure5}
\end{figure}
To describe the magnetic flux density $B(x)$ in the center conductor of the resonator we start with the classical Bean profile\cite{Bean62}.
During an upsweep from the virgin state, the flux density decreases linearly from the edges of the strip and can be expressed as
\begin{equation}
B^{\Uparrow}(x, B_{\textrm{ext}})=
\begin{cases}
\frac{2B^*}{S}|x|-(B^*-B_{\textrm{ext}}), & \frac{S}{2}\geq|x|>\frac{S}{2}b, \\
0, & |x|\leq\frac{S}{2}b, 
\end{cases}
\label{Eq:ProfileBeanUp}
\end{equation}
where $b=(1-\frac{B_{\textrm{ext}}}{B^*})$ and $B^*$ represents the applied field, when the flux fronts from both edges of the strip meet at $x=0$.
After the virgin upsweep to $B_{\textrm{max}}$, the flux profile for the downsweep is given by
\begin{equation}
B^{\downarrow}(x, B_{\textrm{ext}})=B^{\Uparrow}(x, B_{\textrm{max}})-2B^{\Uparrow}(x, \frac{B_{\rm max}-B_{\textrm{ext}}}{2}).
\label{Eq:ProfileBeanDown}
\end{equation}
The flux density profiles $B^{\Uparrow}(x, B_{\textrm{ext}})$ and $B^{\downarrow}(x, B_{\textrm{ext}})$ are schematically shown in Fig.~\ref{fig:Figure5}~(a) and (b) for several values of $B_{\rm ext}$ during (a) a field upsweep to $B_{\textrm{max}}=B^*$  and (b) during the downsweep from $B_{\textrm{ext}}=B^*$ to $B_{\textrm{ext}}=-B^*$.
Note, for each applied magnetic field $B_{\rm ext}$, $|B^{\Uparrow}(x)|\le|B_{\rm ext}|$ and $|B^{\downarrow}(x)|\ge|B^{\Uparrow}(x)|$.
The aforementioned relation also reflects on the dissipation, such that for the same values of applied magnetic field the losses during the downsweep should be larger than in the upsweep.
The lower part of Fig.~\ref{fig:Figure5} shows the calculated quantity $\delta e/\delta e_{\textrm{max}}$ for the Bean model flux profile with a homogeneous ($\lambda_{\textrm{eff}}/S>1$, (c)) as well as a highly inhomogeneous ($\lambda_{\textrm{eff}}/S=0.002$, (d)) microwave current distribution.
All calculations of $\delta e(B)$ were carried out numerically with a spatial resolution of $\Delta x=2\,$nm.
We repeated our calculations with different spacings ($1\,{\rm nm}\le\Delta x\le20\,{\rm nm}$) and found relative deviations $<3\%$.
In case of a homogeneous microwave current density, the dissipation only depends on the number of vortices in the sample, not on their spatial distribution, cf. Eq.~(\ref{Eq:deltae}).
Starting in the virgin state, the total amount of flux increases quadratically with applied field.
Therefore, $\delta e^{\Uparrow}(B)$ also has a positive curvature.
As expected for this field profile (see above), $\delta e^{\downarrow}>\delta e^{\Uparrow}$ and the minimum value of $\delta e^{\downarrow}$, which corresponds to the least amount of flux in the sample, is reached for a negative value of applied field.
In case of a highly inhomogeneous current distribution [Fig.~\ref{fig:Figure5} (d)] only a small region near the edges of the center conductor is responsible for almost all of the dissipation, cf. Fig.~\ref{fig:Figure4}(b).
The flux density in this area is almost identical to $B_{\rm ext}$, hence the hysteresis is much smaller than in the homogeneous case.
Although the hysteresis loop is significantly smaller, certain characteristic features of $\delta e(B)$ remain unchanged, such as the positive curvature of $\delta e^{\Uparrow}(B)$, the relation $\delta e^{\downarrow}>\delta e^{\Uparrow}$ and the fact, that the position of the downsweep minimum is in the negative field range.
Consequently, the classical Bean field profile does not lead to a hysteresis as observed in our experiments.
An alternative model of the flux density in thin film geometries was first considered by Norris~\cite{Norris70} and later discussed by Brandt and Indenbom~\cite{Brandt93}.
It is basically the aforementioned Bean model adapted to the geometry of thin superconducting strips.
In this Norris-Brandt-Indenbom (NBI) model the flux density is given by
\begin{equation}
B^{\Uparrow}(x, B_{\textrm{ext}})=B_c
\begin{cases}
\tanh^{-1}\frac{\sqrt{(x-S')(x+S')}}{|x|\tanh(B_{\textrm{ext}}/B_c)}, & \frac{S}{2}\geq|x|>S' \\
0, & |x|\leq S'
\end{cases}
\label{Eq:NBIm}
\end{equation}
where $S'=S/2\cosh(B_{\textrm{ext}}/B_c)$, $B_c=\mu_0 d j_c/\pi$ is the characteristic field, cf. Ref.~\cite{Brandt93} and $j_c$ is the critical current density of the superconductor.
Using the estimated $j_c\approx 5\times10^{10}\,{\rm A/cm^2}$ of our Nb, we calculate a $B_c\approx 6\,{\rm mT}$. 
As can be seen from Eq.~(\ref{Eq:NBIm}), the NBI model leads to an excess flux density at the edges of the strip compared to the classical Bean profile.
The downsweep flux density $B^{\downarrow}(x,B_{\rm ext})$ is again defined according to Eq.~(\ref{Eq:ProfileBeanDown}).
The NBI flux density profile is shown in Fig.~\ref{fig:Figure6}(a) for several values of $B_{\rm ext}$ during a field upsweep to $B_{\textrm{ext}}=3B_c$ and (b) during the downsweep from $B_{\textrm{ext}}=3B_c$ to $B_{\textrm{ext}}=-3B_c$.
As can be seen in Fig.~\ref{fig:Figure6}~(b), $B^{\downarrow}$ shows a remarkable behavior. For an externally applied field that is still positive, the flux density close to the edge of the thin film is already zero.
When $B_{\rm ext}$ is decreased further, $B^{\downarrow}$ at the film edge becomes negative and the point of zero flux density moves deeper into the sample, separating areas with anti-vortices from that with vortices.
\begin{figure}
\centering {\scalebox{1}[1]{\includegraphics{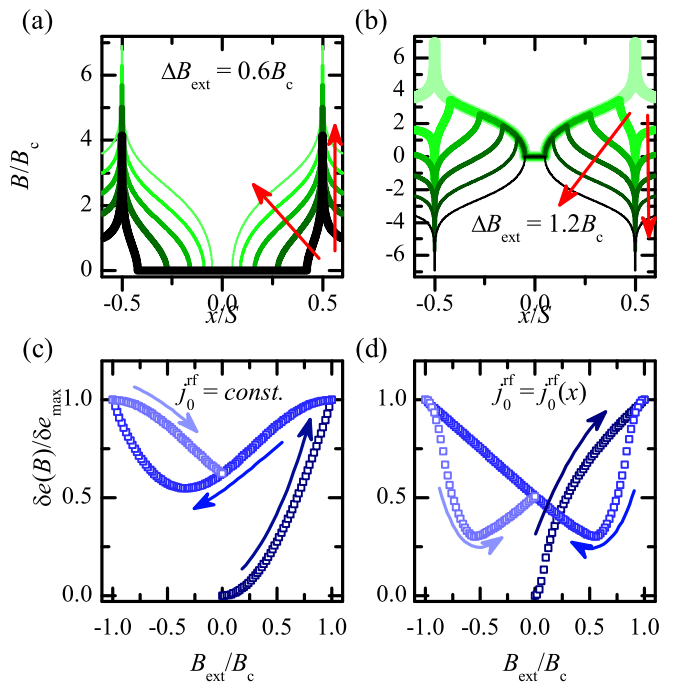}}}
\caption{(Color online) NBI model flux density $B/B_c$ in a superconducting strip of width $S$ during (a) the upsweep to $B_{\textrm{ext}}=3B_c$ ($B_{\textrm{ext}}/B_c=0.6, 1.2, 1.8, 2.4$ and $3$) and (b) during downsweep from $B_{\textrm{ext}}=3B_c$ to $B_{\textrm{ext}}=-3B_c$ ($B_{\textrm{ext}}/B_c=3, 1.8, 0.6, -0.6, -1.8$ and $-3$); Calculated $\delta e(B)/\delta e_{\textrm{max}}$ during a magnetic field cycle $B_{\textrm{ext}}/B_c=0 \rightarrow B_{\textrm{ext}}/B_c=1 \rightarrow B_{\textrm{ext}}/B_c=-1 \rightarrow B_{\textrm{ext}}/B_c=0$ assuming the NBI flux density and (c) a homogeneous as well as (d) a highly inhomogeneous microwave current density with $\lambda_L/S=0.002$ and $d/S=0.006$. Arrows in (a, b) indicate the progression of the flux profile, arrows in (c, d) indicate the sweep direction.}
\label{fig:Figure6}
\end{figure}
To achieve comparability with experimental results, where $B_{\rm max}=4\,{\rm mT}$, a field range of $|B_{\rm ext}|\lesssim B_c\approx 6\,{\rm mT}$ was chosen for numerical calculations.
Figures~\ref{fig:Figure6}~(c,d) show $\delta e(B)/\delta e_{\textrm{max}}$ for a (c) homogeneous and a (d) highly inhomogeneous current density.
In the calculations we avoided the divergence of the NBI flux density at the strip edges by positioning them between two integration points, \ie by effectively introducing a cutoff for $B$ at $\Delta x/2$ from the conductor edges.
In case of $j^{\textrm{rf}}_0(x)=const.$, the NBI and the classical Bean model lead to similar $\delta e(B)/\delta e_{\textrm{max}}$ dependences, cf. Fig.~\ref{fig:Figure5}~(c), which --- as already mentioned --- disagree with our experimental data.
For the inhomogeneous rf current distribution, however, $\delta e(B)/\delta e_{\textrm{max}}$ reproduces almost all characteristic features of the measured curve, cf. Figs.~\ref{fig:Figure3} (a) and ~\ref{fig:Figure6} (d). 
In particular, the hysteresis loop has a butterfly like shape, where $\delta e^{\downarrow}(B)<\delta e^{\Uparrow}(B)$ for a considerable range of $B_{\rm ext}$.
Also, $B^{\Uparrow}$ exhibits a predominantly negative curvature and the minimum of $\delta e^{\downarrow}(B)$ can be found at $B_{\rm ext}>0$.
Yet, there is also a difference between experiment and theory.
An asymmetry between the first very slow increase of the losses at small fields and the abrupt decrease immediately after the inversion of the sweep direction is clearly visible in experiment, but not in theory.
Using the models described before, the curvatures $\partial^2 B^{\Uparrow}(x)/\partial B^2_{\rm ext}$ at $B_{\rm ext}=0$ and $\partial^2 B^{\downarrow}(x)/\partial B^2_{\rm ext}$ at $B_{\rm ext}=B_{\rm max}$ are directly linked, cf. Eq.~(\ref{Eq:ProfileBeanDown}).
Therefore, there is always a symmetry between the first slow increase of the losses at the beginning of the upsweep and the slow decrease at the beginning of the downsweep.
We believe that the asymmetry in experiment originates from the small but nonzero lower critical field $B_{\textrm{c1}}$ of the superconductor, which is not included in the theory. The existence of $B_{\textrm{c1}}$ inherently leads to an asymmetry between the upsweep from the virgin state and any following sweep. 
\begin{figure}
\centering {\scalebox{1}[1]{\includegraphics{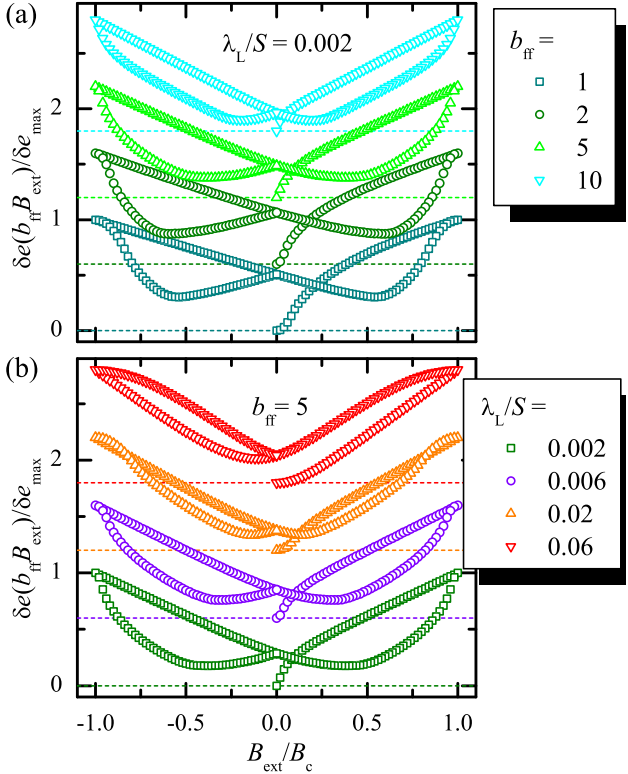}}}
\caption{(Color online) (a) $\delta e(b_{\rm ff}B_{\rm ext})/\delta e_{\textrm{max}}$ vs external flux density $B_{\textrm{ext}}/B_{\textrm{c}}$ calculated for the NBI model with $d/S=0.006$ and different flux focusing factors $b_{\textrm{ff}}$. (b) $\delta e(b_{\rm ff}B_{\rm ext})/\delta e_{\textrm{max}}$ vs. cycled flux density $B_{\textrm{ext}}/B_{\textrm{c}}$ calculated for the NBI model with $d/S=0.006$ and different ratios $\lambda_{\textrm{L}}/S$. Adjacent curves are subsequently shifted by $+0.6$ for better visibility.}
\label{fig:Figure7}
\end{figure}
In Fig.~\ref{fig:Figure7} the dependence of the hysteresis loop on the ratio $\lambda_{\textrm{L}}/S$ and the flux focusing factor $b_{\textrm{ff}}=B_{\textrm{eff}}/B_{\textrm{ext}}$ with the effective flux density $B_{\textrm{eff}}$ in the center-conductor-to-groundplane gaps is shown. 
Figure~\ref{fig:Figure7}~(a) shows $\delta e(b_{\rm ff}B_{\rm ext})/\delta e_{\textrm{max}}$ for four different flux focusing factors $b_{\textrm{ff}}=1, 2, 5$ and $10$.
All curves exhibit the same essential features of the hysteresis but with some minor differences.
With increasing flux focusing, the field regions around $B_{\rm ext}=0$ and $B_{\rm ext}=B_{\rm max}$ where curvatures of $\delta e^{\Uparrow}$ and $\delta e^{\downarrow}$ change sign become effectively compressed. 
Furthermore, the position of the downsweep minimum shifts with $b_{\rm ff}$.
By comparing calculations with experimental data, we find best qualitative agreement for a flux focusing factor between  $b_{\textrm{ff}}\approx 2$ and $b_{\textrm{ff}}\approx 5$.
Although the estimated value of $B_c\approx 6\,{\rm mT}$ is comparable to $B_{\rm max}=4\,{\rm mT}$ in experiment and a flux focusing factor of $b_{\textrm{ff}}\approx 5$ is compatible with the resonator geometry, we would like to emphasize, that the theoretical model presented in this paper is too simple to be used for a quantitative analysis of our experimental data.
%
%
%
%

%
Besides the flux focusing factor the homogeneity of the rf current distribution described by $\lambda_L/S$ determines the shape of the hysteresis loop.
Figure~\ref{fig:Figure7}~(b) shows $\delta e(B)/\delta e_{\textrm{max}}$ for four different ratios $\lambda_{\textrm{L}}/S=0.002, 0.006, 0.02$ and $0.06$, the corresponding current distributions are shown in Fig.~\ref{fig:Figure4}~(b).
Starting from $\lambda_{\textrm{L}}/S=0.002$, which corresponds to our experimental conditions, the hysteresis loop becomes smaller with increasing homogeneity.
For $\lambda_{\textrm{L}}/S\approx0.02$ hardly any hysteresis can be seen.
When $\lambda_{\textrm{L}}$ is increased further, up- and downsweep curves do not even cross anymore. 
Due to the good agreement between theory and experimental data, some important implications of our model and $\delta e(B)$ calculations presented in this paper should be addressed. 
In our calculations we have assumed, that the microwave current density and the flux (vortex density), can be treated independently from each other.
In general, however, the presence of a Bean-like flux gradient leads to a redistribution of a transport current and vice versa \cite{Brandt93}.
Also, the oscillation amplitude of the vortices $\Delta x$ is assumed to be much smaller than the length scale on which the current distribution varies, such that $j_0^{\rm rf}(x\pm \Delta x)\approx j_0^{\rm rf}(x)$.
We believe, that both assumptions are reasonable, as in experiment we observe, that the vortex associated losses $1/Q_v$ are almost independent of the power of applied microwaves, as long as $P_{\rm app}<0\,$dBm~\cite{Bothner11}.
If the microwave self-field significantly disturbed and rearranged the static magnetic field configuration by introducing additional vortices, we would expect some nonlinearities \cite{Golosovsky95}, which would lead to a power dependence of the losses.
Further simplifications are, that in experiment, the flux focusing factor depends on the applied flux density, as the external flux is only partially focused into the gaps of the coplanar waveguide.
Some of the flux also penetrates the superconductor in form of Abrikosov vortices.
Moreover, due to the geometry of our samples with a meandered resonator line, $b_{\rm ff}$ is expected to vary along the resonator.
Consequently, also the losses $\delta e=\delta e(z)$ depend on position along the resonator.
Finally we ignore the ground planes and that they might experience a different effective $B$ at their edges and hence a different hysteresis as the center conductor.
In the experiments all these effects are merged together, but we think that none of them fundamentally changes the observed and analyzed hysteretic behavior.

\section{Demagnetization and tunability}

\label{sec:Demagnetization}
In order to explore the possibility to return to the virgin state after magnetic cycling, we performed a demagnetization procedure, \textit{i.e.} we swept the magnetic field up and down again with decreasing amplitude and monitored the corresponding resonances.
The procedure slightly differs from commonly used demagnetization cycles, as we kept the step width of the magnetic field sweep constant and thus with decreasing amplitude of the sweep, the oscillation period between positive and negative maximum became smaller accordingly.
Still, in the following we will refer to this procedure as demagnetization.
\begin{figure}
\centering {\scalebox{1}[1]{\includegraphics{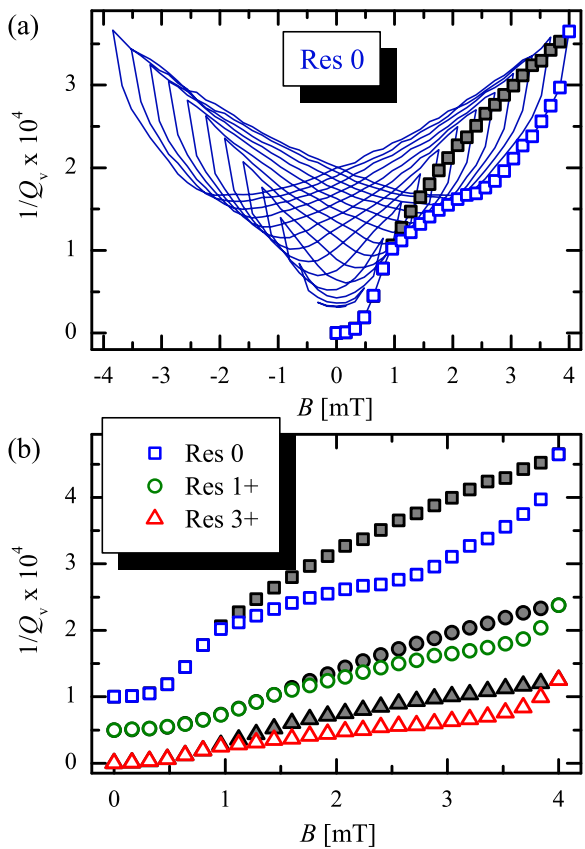}}}
\caption{(Color online) (a) Measured vortex associated energy losses $1/Q_v(B)$ of Res \textbf{0} of the fundamental mode $n=1$ for a full demagnetization cycle (line), for the virgin field sweep (full squares) and minimum values for positive magnetic fields (open squares); (b) $1/Q_v(B)$ of the resonators \textbf{3+} (triangles), \textbf{1+} (circles, shifted by $+5\times10^{-5}$) and \textbf{0} (squares, shifted by $+1\times10^{-4}$) for the virgin field sweep (full symbols) and the minimum values for each magnetic field values during the demagnetization sweep (open symbols).}
\label{fig:Figure8}
\end{figure}
Figure~\ref{fig:Figure8} (a) shows the measured energy losses $1/Q_v(B)$ during a full demagnetization procedure (blue line).
As can be seen, with decreasing amplitude of the up- and downsweeps, the losses at zero magnetic field also decrease and approach, yet do not fully return, to the virgin state value. At the end of the demagnetization a small number of vortices remains in the resonator.
During each second half cycle Abrikosov anti-vortices (``anti'' relative to the remanent ones from the first half cycle) are pushed into the sample almost as far as the remanent vortex front reaches.
The remaining losses after demagnetization probably reflect remanent vortices (and possibly anti-vortices as well) which were not annihilated during the repeated cycle.
To avoid non-equilibrium effects and to ensure an adiabatic magnetic field sweep, we waited about one minute after each field step. 
The demagnetization curve in Fig.~\ref{fig:Figure8} (a) shows, that over almost the whole magnetic field range the envelope of the full curve, \ie the minimum values of the vortex associated losses (open squares), lies considerably below the values of the virgin field sweep (full squares).
This provides the opportunity to significantly reduce the losses of resonators or other microwave circuitry components operated in magnetic fields (e.g. when it comes to trapping of an ultracold atom cloud in the vicinity of a resonator) by proper choice of magnetic history.
We are aware, that the achievable reduction is small compared to other approaches such as using antidots or slots, but nevertheless our procedure can be used to further improve the device performance by about $30\%$.
\begin{figure}
\centering {\scalebox{1}[1]{\includegraphics{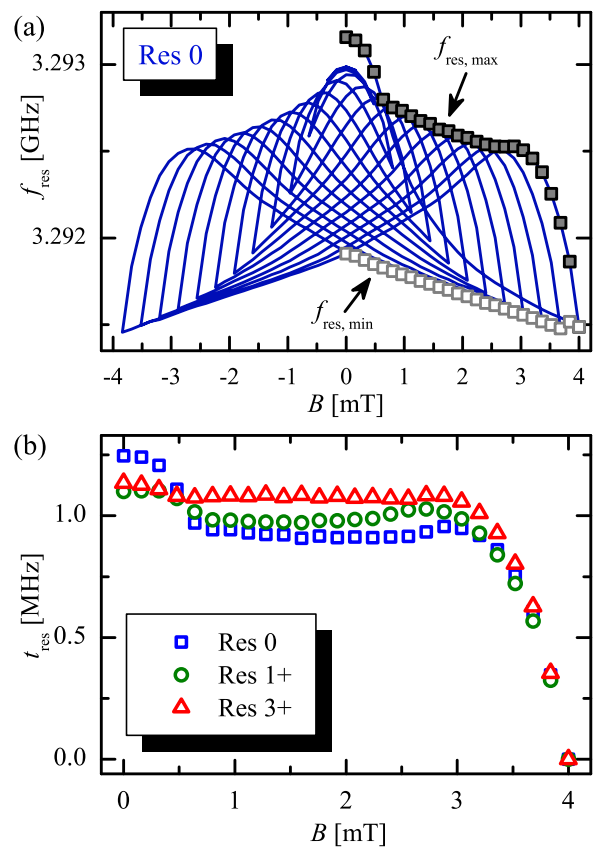}}}
\caption{(Color online) (a) Measured resonance frequency $f_{\textrm{res}}(B)$ of Res \textbf{0} for a full demagnetization cycle (line) with maximum values (full squares) and minimum values (open squares) for positive magnetic fields; (b) Tunability $t_{\textrm{res}}(B)=f_{\textrm{res, max}}(B)-f_{\textrm{res, min}}(B)$ of the resonators \textbf{3+} (triangles), \textbf{1+} (circles) and \textbf{0} (squares).}
\label{fig:Figure9}
\end{figure}
The demagnetization cycle behavior of the resonance frequency $f_{\textrm{res}}$, is depicted in Fig.~\ref{fig:Figure9} (a).
The curve strongly modulates and covers a range of about $1\,$MHz for all magnetic field values.
For a quantification, we define the frequency tunability of the resonator as the difference between the maximum and the minimum resonance frequency $t_{\textrm{res}}\equiv f_{\textrm{res, max}}-f_{\textrm{res, min}}$ for each value of applied magnetic field.
The resulting $t_{\textrm{res}}(B)$ of the three resonators with and without antidots is shown in Fig.~\ref{fig:Figure9} (b) for positive values of magnetic field.
For all three resonators the tunability is about $t_{\textrm{res}}(B)=1\,$MHz and hence seems to be almost independent of a perforation and magnetic field with a small tendency to increase with the number of antidots.
A tunability like this might be useful, e.g. when it comes to a finetuning of the superconducting cavity to the (fixed) transition frequencies of ultracold atom clouds.
As each frequency within the tunability range at each field point is accessible by at least two different histories and the maxima/minima of the frequency/loss hysteresis are not at the same field values, one has to check all possibilities to find the optimal combination of desired frequency and losses.
As the exact parameters of the hysteresis slightly differ from device to device and probably from setup to setup, we can not give a common recipe for finding the best working point here, but each device has to be pre-characterized in the corresponding experimental situation.
We emphasize, that we do not propose to tune the properties of planar superconducting microwave components by just applying a magnetic field and introducing Abrikosov vortices here.
If one just would like to fabricate a frequency tunable device for instance, other approaches with larger ranges and tunability velocities seem more promising \cite {Healey08, Palacios08, Sandberg08}.
But if the microwave components have to be operated in specific magnetic fields anyway, magnetic history effects provide a nice additional possibility to reduce the losses and tune the resonance frequencies by spatially re-arranging Abrikosov vortices.

\section{Higher modes and other angles}

\label{sec:Angles}
\begin{figure}
\centering {\scalebox{1}[1]{\includegraphics{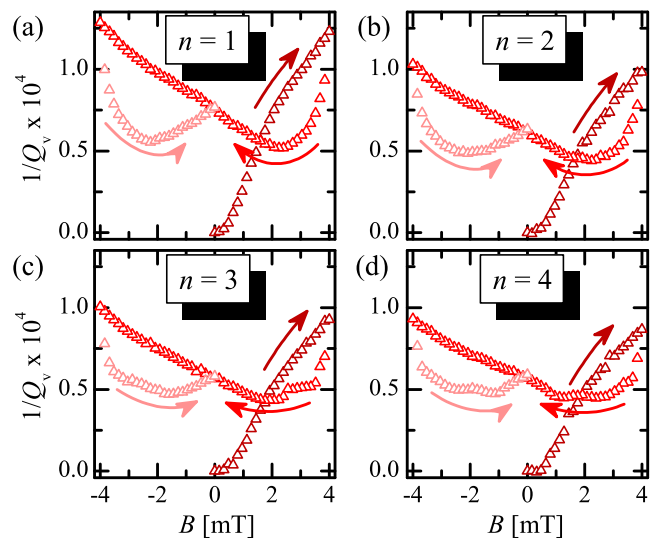}}}
\caption{(Color online) Measured vortex associated energy loss $1/Q_v(B)$ for the four lowest modes $n=1$, $n=2$, $n=3$, $n=4$ of Res \textbf{3+} for a full cycle of magnetic field.}
\label{fig:Figure10}
\end{figure}

\begin{figure}	
\centering {\scalebox{1}[1]{\includegraphics{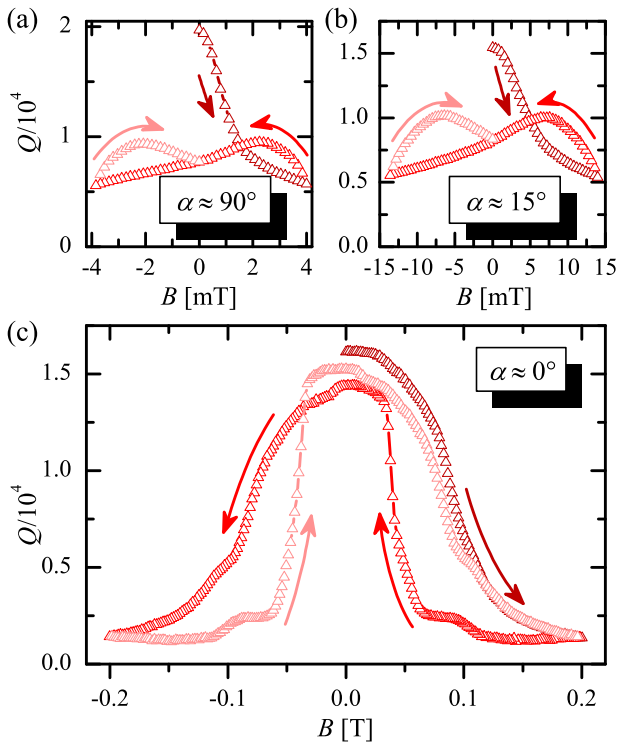}}}
\caption{(Color online) Typical measured hysteresis curves of the quality factor $Q$ of the fundamental mode $n=1$ (resonator 3+) during a full cycle of the magnetic field for three different angles (a) $\alpha=90^{\circ}$, (b) $\alpha\approx 15^{\circ}$ and (c) $\alpha\approx 0^{\circ}$ between resonator plane and applied field.}
\label{fig:Figure11}
\end{figure}

So far we have only considered the fundamental mode of our resonators and an operation in a perpendicular magnetic field.
In the following we will present measurement results on higher modes and discuss the influence of the tilt angle between resonator plane and magnetic field on the hysteresis loop.
First, in Fig.~\ref{fig:Figure10} we exemplarily show the measured energy losses $1/Q_v(B)$ of resonator \textbf{3+} for (a) the fundamental frequency $n=1$ and (b), (c), (d) the first three harmonics $n=2$, $n=3$, $n=4$ in perpendicular magnetic fields.
Clearly, the characteristic features of the hysteresis loop can also be seen for higher modes.
Interestingly, with increasing frequency a fine structure within the hysteresis curves emerges, see \eg Fig.~\ref{fig:Figure10}~(d) at $B\approx\pm1.5\,{\rm mT}$.
This fine structure might be related to inhomogeneously distributed losses along the resonator. 
Depending on the resonator mode $n$, nodes and anti-nodes of the microwave current standing wave probe different parts of the resonator, \ie they contribute differently to the overall losses. 
As already pointed out, there is a variation of the flux focusing factor and thus the effective flux density along the resonator due to its geometry.
The flux focusing is smaller in the meandered lines than in the straight part at the midpoint of the resonator, as the adjacent gaps of the meander allow for partial sharing of flux.
Consequently, for different flux focusing factors the downsweep minima might occur at different applied magnetic fields and the superposition of the $1/Q_v$ curves of several antinodes (which is, what we measure) might lead to a fine structure on the hysteresis loop with a multiple minima.
We also performed measurements, where the magnetic field was not applied perpendicular to the resonator plane but with an angle $\alpha\neq90^{\circ}$ (cf. Fig.~\ref{fig:Figure4}~[a]).
To do so, we used a different setup, where we rotated the sample in the field of a stationary superconducting high-field split coil.
By rotating the sample and measuring the field dependent quality factor for different angles we determined the perpendicular and parallel orientations between magnetic field and resonator, assuming that for a given magnetic field losses are maximal for perpendicular and minimal for parallel orientation.
Note, a comparison of the measured losses for the perpendicular orientations in the two setups showed, that they are practically identical, although the zero field quality factor is smaller in the high-field setup probably due to a remanent magnetization of the cryostat components.
Figure~\ref{fig:Figure11} shows the measured dependence of the quality factor $Q$ on the applied field $B$ of the fundamental mode of resonator \textbf{3+} for three different angles.
In perpendicular fields, $\alpha=90^{\circ}$, we find the well-known and already described hysteresis with significantly reduced losses $1/Q_v$ (increased quality $Q$) on the downsweep branch, see Fig.~\ref{fig:Figure11}~(a).
For $\alpha<90^{\circ}$, the hysteresis loop hardly changes, as can exemplarily be seen in Fig.~\ref{fig:Figure11}~(b), where $\alpha \approx 15^{\circ}$.
Note, here the field range is about four times larger compared to the measurement with perpendicular orientation (a) and the general shape of the hysteresis and the downsweep improvement in $Q$ are clearly visible.
As $\sin (15^{\circ})\approx 0.26$, measurement results strongly suggest, that the resonator losses are primarily determined by the component of $B$ perpendicular to the sample.
The situation changes dramatically for $\alpha\approx0^{\circ}$, where the magnetic field was swept between $\pm 0.2\,$T, as can be seen in Fig.~\ref{fig:Figure11} (c).
For magnetic flux densities $B<80\,$mT the decrease in $Q$ is relatively small and also reversible (not shown).
When the magnetic field is increased further, the quality factor rapidly decreases. 
After sweep direction inversion $Q$ remains low until $B\lesssim 50\,$mT, when the quality factor increases and almost completely recovers to the original value of $Q(0)$.
For negative fields we observe qualitatively the same behavior.
It is important to note, that $Q^{\downarrow} \leq Q^{\Uparrow}$.
Interestingly, a comparison of measurement results with the theoretical curves presented in Sec.~\ref{sec:Model} shows best agreement with the prediction of the classical Bean model and not the NBI model, cf. Fig.~\ref{fig:Figure5}.
Although on the first view one indeed would expect the classical Bean model to be the adequate description for the flux density profile in this experimental situation (field parallel to a superconducting plane), one has to be careful with this interpretation for two reasons.
First one has to consider, that the film thickness $d=300\,$nm is about three times the penetration depth ($\lambda_{\textrm{eff}}\approx100\,$nm), and therefore only $1.5$ times the diameter of one Abrikosov vortex.
Hence the Bean model as kind of mean field theory might probably need a modification to be suitable for calculating the resonator losses in this case.
The step-like structures in $Q(B)$ (\eg at $B\approx\pm0.1\,$T) might reflect some kind of a discrete and non-smooth flux entry into and exit out of the superconductor.
Moreover, it is likely, that the measured hysteresis is also partly the result of a small misalignment between the applied field and the resonator plane. 
At flux densities of $B\approx 100\,$mT, an alignment error of only $0.5^{\circ}$ already introduces a field of $\approx 1\,$mT perpendicular to the film, which is large enough to significantly reduce $Q$, cf. Fig.~\ref{fig:Figure11}~(a).
It remains to be stated at the end of this discussion, that the hysteresis loop seems not to allow for reducing the microwave losses in the resonator, if the magnetic field is applied close to parallel to the superconducting film.

\section{Conclusions}

\label{sec:Conclusion}
We experimentally investigated the properties of $3.3\,$GHz superconducting coplanar transmission line resonators in magnetic fields.
We particularly focused on resonators, that were zero field cooled to $T=4.2\,$K and afterwards exposed to magnetic fields cycled in the milli-Tesla range.
We measured the resonance frequency $f_{\textrm{res}}$ and quality factor $Q$ of the resonators and found strong hysteresis effects, which on the mesoscopic scale are due to presence of Abrikosov vortices in the superconducting film and their spatial redistribution during the field cycles.
Using a simple model for the vortex associated resonator losses we show that different combinations of microwave current and flux density distributions lead to characteristically different hysteresis loops. 
We find best agreement between experiment and theory for a current distribution strongly peaked at the resonator edges and a modified Bean flux gradient for thin films, as described by Norris, Brandt and Indenbom.
We have also shown, that the hysteresis may be used to improve the resonator performance for fixed values of applied magnetic field by proper choice of magnetic history.
Accordingly, the resonance frequency can be tuned by about $1\,$MHz, \textit{i.e.} for our resonators a few zero-field linewidths at liquid Helium temperature.
Both, the reduction of the losses and the tunability of the frequency are found to be qualitatively and quantitatively independent of a pre-patterning of the resonators with antidots, which in a previous study has been shown to be feasible to reduce vortex associated losses.
Furthermore, we show, that the hysteresis can also be found for higher modes $n=2, 3, 4$ of the resonators and for angles between $90^{\circ}$ and a few degrees between the field direction and the resonator plane.
In the parallel orientation, the hysteresis showed a very different behaviour with presumably no possibility for a resonator improvement with magnetic history.
As for many experiments in circuit quantum electrodynamics the resonators are operated in the Millikelvin and single photon regime, the effects presented here have to be investigated under these conditions in further studies, but there are definitely no obvious reasons, why the reported hysteresis effects should qualitatively change with decreasing temperature and power.
Other parameters, which are expected to change the nature of the hysteresis and have to be considered for possible applications are the geometric dimensions of the transmission line and the thickness of the films, which together with the magnetic penetration depth have a strong influence on current distribution and flux density profile.

\section{Acknowledgements}

\label{sec:Acknowledgements}
This work has been supported by the Deutsche Forschungsgemeinschaft via the SFB/TRR 21 and by the European Research Council via SOCATHES.
It was partly funded by the German Federal Ministry of Education and Research (Grant No. 01BQ1061).
DB acknowledges support by the Evangelisches Studienwerk e.V. Villigst.
MK acknowledges support by the Carl-Zeiss Stiftung.
%
%
The authors thank Roger W\"ordenweber for fruitful discussions.

\end{document}